%% file: internettrustv18.tex
\documentclass[12pt,english]{article}
\usepackage[OT1]{fontenc}
\usepackage[utf8]{inputenc}
\usepackage[english]{babel}
\usepackage{float}
\usepackage{amsmath}
\usepackage{amsthm}
\usepackage{amssymb}
\usepackage{graphicx}
\usepackage{setspace}
\usepackage{esint}
\usepackage[authoryear]{natbib}
\usepackage[unicode=true,pdfusetitle,
 bookmarks=true,bookmarksnumbered=false,bookmarksopen=false,
 breaklinks=false,pdfborder={0 0 1},colorlinks=false]
 {hyperref}
\hypersetup{
 urlcolor=blue}
\usepackage{breakurl}
\usepackage[a4paper, text={6.2in,9.2in}, centering]{geometry}
\usepackage{abstract}
\usepackage{array}
\usepackage{varioref}
\usepackage{caption}
\usepackage[table]{xcolor}
\usepackage{booktabs}
\usepackage{longtable} 
\usepackage{threeparttable}
\usepackage{rotfloat} 
\usepackage{graphicx}
\usepackage{placeins}
\usepackage{float}
\usepackage{subfigure}
\usepackage[toc,page]{appendix}

\begin{document}

\title{Broadband Internet and Social Capital}

\author{Andrea Geraci\thanks{European Commission, Joint Research Centre (JRC); University of Oxford,
INET. \protect\\
andrea.geraci@ec.europa.eu}, Mattia Nardotto\thanks{KU Leuven, Department of Management Strategy \& Innovation; CEPR;
CESifo. \protect\\
mattia.nardotto@kuleuven.be}, Tommaso Reggiani\thanks{Masaryk University, Department of Public Economics; IZA. \protect\\
tommaso.reggiani@econ.muni.cz}, Fabio Sabatini\thanks{Sapienza University of Rome, Department of Economics and Law. \protect\\
fabio.sabatini@uniroma1.it } }
\maketitle
\begin{abstract}
\noindent We study how the diffusion of broadband Internet affects
social capital using two data sets from the UK. Our empirical
strategy exploits the fact that broadband access has long depended
on customers' position in the voice telecommunication infrastructure
that was designed in the 1930s. The actual speed of an Internet 
connection, in fact, rapidly decays with the distance of the dwelling
from the specific node of the network serving its area. Merging unique
information about the topology of the voice network with geocoded
longitudinal data about individual social capital, we show that access
to broadband Internet caused a significant decline in forms
of offline interaction and civic engagement. Overall, our results
suggest that broadband penetration substantially crowded out several
aspects of social capital. \bigskip{}

\noindent \textbf{Keywords}: ICT, broadband infrastructure, networks,
Internet, social capital, civic capital.

\noindent \textbf{JEL Classification}: C91, D9, D91, Z1.\newpage{} 
\end{abstract}

\clearpage

\section{Introduction}

Social capital has been credited with playing a role in many desirable
outcomes such as access to credit and loan repayment (\citealp{Karlan_2005_AER};
\citealp{Feigenberg_et_al_2013_RES}), financial development (\citealp{Guiso_Sapienza_Zingales_2004_AER}),
innovation (\citealp{Knack_Keefer_1997_QJE}), mitigation of agency
problems in organizations (\citealp{Costa_Kahn_2003_QJE}), political
accountability (\citealp{Nannicini_et_al_2013}), and economic growth
(\citealp{Algan_Cahuc_2010}; \citealp{Guiso_2016_JEEA}), just to
name a few. In his best seller \textit{Bowling Alone}, \citet{Putnam_2000_Bowling_Alone}
documented that a decline in measures of social capital–such as participation
in formal organizations, informal social connectedness, and interpersonal
trust–began in the United States in the 1960s and 1970s, with a sharp
acceleration in the 1990s. This study prompted many subsequent empirical
tests that mostly supported Putnam's ``decline of community life
thesis” (\citealp{Paxton_1999_AJS}). \citet{Costa_Kahn_2003_Kyklos},
for example, found a decline in indicators of volunteering, membership
of organizations and entertainment with friends in the United States
since 1952. \citet{Li_et_al_2003_BJS} recorded similar patterns for
the UK starting from 1972.

In \textit{Bowling Alone}, three explanations were put forward for
the decline in social capital: a) the reduction in the time available
for social interaction—related to the rise in labor flexibility and
to the expansion in commuting time; b) the rise in mobility of workers
and students; c) progress in information and communications technology (ICT). \citet{Putnam_2000_Bowling_Alone}
argued that television and other forms of domestic entertainment such
as video games and video players displaced relational activities in
individuals' leisure time. This argument is consistent with empirical
evidence of a negative role of television exposure in social relations
(\citealp{Bruni_Stanc_2008_JEBO}), civic engagement (\citealp{Olken_2009_AEJ_AP}),
and voter turnout (\citealp{Gentzkow_2006_QJE}). As television, a
unidirectional mass medium, was found to significantly affect aspects
of social capital, it stands to reason that the Internet, which provides on-demand content and allows for interactive communication, might induce
an even more powerful substitution effect.

Given the pervasiveness of the Internet and the importance of social
capital in economic outcomes, the effect of broadband penetration
should be put under scrutiny by economic research. Studying the relationship
between high-speed Internet and social capital, however, poses several
methodological problems. The available longitudinal surveys contain
limited information regarding Internet access and use. We lack longitudinal
data about the time interviewees spend using the Internet and their
online activities. 
In addition, the use
of survey data entails endogenous sample selection and 
treatment assignment. As a result, it is difficult to establish causality. The
purchase of a fast Internet connection and aspects of social capital
such as interpersonal interactions and civic engagement may be co-determined
by unobservable personality traits. Reverse causality is also at stake,
as more socially active individuals may have a stronger propensity
for using the Internet as a tool to preserve and extend their offline relationships.
As a result, the existing evidence on the role of the Internet in the accumulation
of social capital is limited and mostly anecdotal.

This paper studies the effect of the introduction of high-speed Internet
on social capital using two data sets from the UK. To
address endogeneity concerns, we exploit exogenous discontinuities
in the quality of Internet access. During the time period considered
in this paper, access to fast Internet in the UK was mostly
based on the digital subscriber line (DSL) infrastructure. DSL technology
allows the high-speed transmission of data over the old copper telephone
network. However, the existence of a voice network is a necessary
but not a sufficient condition for the availability of broadband.
The actual speed of a domestic connection rapidly decays with the
distance of a final user's telephone line from the node of
the network serving the area, also called local exchange (hereafter LE). %
Our empirical analysis exploits the fact that, while at the time the
network was designed in the 1930s the length of the copper wire connecting
houses to the LE did not affect the quality of voice communications,
the introduction of DSL technology in the 1990s unpredictably turned
distance from the LE into a key determinant of Internet access and
quality. Proximity to the relative node of the network thus resulted
in access to fast Internet, while more distant dwellings were de facto
excluded from accessing the broadband.\footnote{\citet{Ofcom2011}.} This identification strategy is similar to those developed in recent
studies that exploit distance from the LE (or between the LE and higher
nodes of the network) to identify the causal effect of broadband Internet
on political participation (\citealp{Falck_Gold_Heblich_2014_AER};
\citealp{Campante_2017_JEEA}), electoral results (\citealp{Miner_2015_JPUBE}), health (\citealp{Amaral_et_al_2017_health}),
and fertility (\citealp{Billari_et_al_2017_IZA}).

Merging unique information about the topology of the voice network provided by Ofcom 
- including the geolocation of LEs and of the city blocks served by
each of them - with geocoded individual survey data
from the British Household Panel (BHPS) over the period 1997-2009,
we are able to assess how broadband access affected several dimensions
of social capital, such as participation in voluntary organizations, political participation, the frequency of voluntary work, and certain types of cultural consumption
that are usually enjoyed in company such as cinema attendance.

First, we use our detailed map of the topology of the network to calculate
the distance of the LE serving each Lower Layer Super Output Area
(LSOA) from the centroid of the area. LSOAs are the second-narrowest
geographical areas in the UK census, comprising on average 650 households
and 1,500 inhabitants. In densely populated metropolitan areas they
correspond to portions of city blocks. We then match this information with the geographic
coordinates of the households surveyed in the BHPS. Since the availability of fast Internet was strongly affected, for technological reasons, by the physical distance between the LE and the premises, we are able
to employ an intention-to-treat approach to assess the effect of fast
Internet on social capital.

We find that after the advent of the broadband in the area, several
indicators of civic engagement and offline interaction started to
decrease with proximity to the node of the network, suggesting that
the exposure to fast Internet displaced social capital.
Placebo tests support the causal interpretation of our results. The panel
structure of the dataset allows us to confirm that distance
from the LE is not associated with changes in individual social capital
before the penetration of the broadband.

Our paper bridges three strands of literature. The first broadly includes
empirical research on the sources of social capital. Most studies
in this field investigate the persistent role of exogenous variations
that took place decades or centuries ago, such as inherited culture
(\citealp{Algan_Cahuc_2010}), long-gone formal institutions (\citealp{Becker_et_al_2014_EJ}),
a community's past history of independence (\citealp{Guiso_2016_JEEA}),
and slave trade (\citealp{Nunn_Wantchekon_2011_AER}). We add to this
field by investigating how progress in ICT can induce a much more
rapid, though presumably persistent, change in the stock of social
capital. Our paper is particularly related to contributions assessing
the response of social capital to other contingent stimuli such as
conflict (\citealp{Guriev_Melnikov_2016_AER}), exposure to television
(\citealp{Bruni_Stanc_2008_JEBO}), teaching practices (\citealp{Algan_Cahuc_Shleifer_2013_AEJ_AP}),
unemployment (\citealp{Algan_et_al_2017_BPEA}), and regulation (\citealp{Aghion_et_al_2010_QJE}).

The second strand encompasses sociological
and economic studies that analyze how social engagement may be substituted by private activities (such as videogames) as new technology-intensive devices become more accessible. Several sociologists suggest that Internet use may displace forms of social capital in individuals' leisure time (\citealp{Kraut_et_al_1998_APSY};
\citealp{DiMaggio_et_al_2001_ARS}; \citealp{Wellman_Hampton_2001_ABS}).
\citet{Antoci_et_al_2011_JEBO} developed a dynamic model to analyze
how technological progress can lead individuals to replace relational
goods with material consumption. This substitution process may trigger
a chain of reactions leading society into a ``social poverty trap''
(\citealp{Antoci_et_al_2015_JMAS}). We add to this field by offering
an empirical test of the possible displacement effect caused by progress
in ICT.

The third strand studies the effect of domestic broadband access or
penetration on political participation (\citealp{Czernich_2012_KYKL};
\citealp{Falck_Gold_Heblich_2014_AER}; \citealp{Campante_2017_JEEA}), electoral results (\citealp{Miner_2015_JPUBE};
\citealp{Gavazza_Nardotto_Valletti_2018_RES}), economic growth (\citealp{Czernich_et_al_2011_EJ}),
social capital (\citealp{Bauernschuster_2014_JPUBE}), fertility (\citealp{Billari_et_al_2017_IZA}),
sex crime (\citealp{Bhuller_et_al_2013_RES}; \citealp{Nolte2017}), health (\citealp{Amaral_et_al_2017_health}),
and well-being (\citealp{Castellacci_Tveito_2018_RP}). Recent studies also focus on the role of the Internet 2.0, highlighting the potential of social media to support 
collective action and political mobilization in young democracies
or authoritarian regimes (\citealp{Enikolopv_et_al_2016_CEPR}; 
\citealp{Enikolopv_et_al_2016_AEJ}), but also to exacerbating polarization
(\citealp[a]{Muller_Schwartz_2018_SSRN}; \citeyear[b]{Muller_Schwartz_2018_hate_}),
foster the spread of misinformation (\citealp{DelVicario_et_al_2016_PNAS}), and
biasing electoral results (\citealp{Allcot_Gentzkow_2017_JEP}) in older
democracies.
In the paper that is closer in spirit to ours, \citet{Bauernschuster_2014_JPUBE}
exploited a quasi-experiment created by a technology choice of the
state-owner provider, which unintentionally hindered broadband penetration
in some areas, to identify the effect of fast Internet on social capital
in East Germany. We differentiate from this study in several respects.
Our identification strategy, though similar to that in \citet{Bauernschuster_2014_JPUBE},
exploits unique information about the distance between customers'
dwellings and the nodes of the voice network. We use an intention-to-treat
approach to test whether changes in the social capital of households
occurred after the penetration of the broadband in relation to users'
position in the topology of the network resulting in the actual access
to fast Internet. The analysis was conducted in the UK
over a period of 10 years. Finally, we assess the effect of broadband
Internet on a wider range of indicators encompassing both the structural
and the cognitive dimensions of social capital. 

The remainder of this paper is organized as follows. Section 2 
briefly reviews the literature on the effect of Internet-mediated communication
on social capital. Section 3 describes the diffusion
of broadband Internet in the UK. Section 4 describes the
data. Section 5 presents our our empirical analysis of the effect
of broadband Internet on social capital. Section 6 discusses 
how the topology of the network affected the Internet take up
Section 7 concludes.

\section{Social capital and the Internet }

Social capital is generally referred to as all features of social
life – networks, norms, civic engagement and trust – that enable individuals
to act together more effectively to pursue shared objectives (\citealp{Putnam_1995_JD}).
The literature has provided so many definitions of social capital
that clarifying the dimensions of the concept has long been a research
priority. \citet{Uphoff_1999_inbook} proposed a distinction between
structural and cognitive dimensions: structural social capital refers
to individuals' behaviors and consists of social participation and
civic engagement (e.g. meetings with friends and membership in organizations).
Cognitive social capital derives from individuals' perceptions resulting
in trust, values and beliefs that promote pro-social behavior.

Putnam's (\citeyear{Putnam_2000_Bowling_Alone}) concern about the
detrimental role of technology related to the structural dimension of
social capital. In \textit{Bowling Alone}, the author borrowed two
main arguments from the sociological literature to explain how ICT
development could crowd out face-to-face interaction. First, the more
time people use the Internet for leisure, the more time is
detracted from social activities like communicating with friends,
neighbors and family members (\citealp{DiMaggio_et_al_2001_ARS};
\citealp{Nie_et_al_2002_}; \citealp{Gershuny_2003_SFORCES}; \citealp{Wellman_Hampton_2001_ABS}).
The second argument relies on the concept of ``community without
propinquity'' (\citealp{Webber_1963_inbook}) and on the earlier
theories of the Chicago School of Sociology. In a famous paper, \citet{Wirth_1938_AJS}
claimed that any increase in the heterogeneity of the urban environment
would provoke the cooling-off of ``intimate personal acquaintanceship''
and would result in the ``segmentation of human relations'' into
those that were ``largely anonymous, superficial, and transitory''
(\citealp{Wirth_1938_AJS}, p. 1). This line of reasoning can be easily
applied to the Internet, which has the potential to fragment local
communities into new virtual realities of shared interest that may
negate the necessity of face-to-face encounters (see for example \citealp{Antoci_et_al_2011_JEBO}
and \citealp{ConradsReggiani2017}).

Empirical tests of the crowding out hypothesis have mostly been conducted
in the fields of sociology and psychology and mainly focus on the
structural dimension of social capital. In one of the earliest studies,
\citet{Kraut_et_al_1998_APSY} observed 69 individuals from 93 households
in Pittsburgh for two years, concluding that increased Internet use
was associated with a decline in interactions with family members
within the household, a reduced social circle, and a rise in loneliness
and depression. \citet{Nie_Erbring_2000_report} used U.S. cross-sectional
data to show that Internet users reported spending less time with
family and friends than non-users. \citet{Stepanikova_et_al_2010_CHB}
used panel time-diary data collected in a group of U.S. residents
in 2004 and 2005 to examine whether loneliness and life satisfaction
were associated with time spent on various Internet activities. Cross-sectional
models revealed that time spent using the Internet was positively
related to loneliness. However, the relationship was not robust to
time fixed effects, suggesting that cross-sectional results could
be driven by endogeneity issues such as omitted variables bias and
reverse causality.

The concern for reverse causality is supported by studies finding
that lonely people tend to use the Internet more. Drawing on data
from a field study with 89 participants, \citet{Hamburger_BenHartzi_2003_CHB}
found that lonely people spend more time using the Internet than non-lonely
people. Analyzing responses from a survey of 277 undergraduate Internet
users, \citet{Morahan_Martin_Schumacher_2003_CHB} suggested that
lonely individuals may be drawn online because of the increased potential
for companionship and as a way to mitigate negative moods associated
with loneliness.

This literature mostly refers to the first stage of the Internet penetration
that took place before the advent of social networking sites (SNS).
More recent studies specifically focusing on the role of social media
such as Facebook support the hypothesis that online interaction may
instead help preserve social ties against time and space constraints.
SNS have been found to allow the crystallization of weak or latent
ties which might otherwise remain ephemeral (\citealp{Ellison_2007}),
boost teenagers’ self-esteem by encouraging them to relate to their
peers (\citealp{Steinfield_2008}), and promote interest in politics
and public affairs in general (\citealp{GildeZuniga_et_al_2012_JCMC}).

Overall, this literature points out two opposing trends. While Internet access per se has been
found to displace offline activities, social networking sites seem
to have the potential to facilitate interaction and coordination among
agents. These results are not as conflicting as they may seem. The first
wave of Internet use, 
in fact,
mainly consisted of asynchronous interaction through emails and forums
and of the consumption of on-demand contents such as news sites and blogs. As the time devoted to
web surfing necessarily had to be detracted from other activities
such as face-to-face interaction, early sociological studies explained
the negative correlation between Internet use and  social
participation as a result of a trade-off in the use of time. The web
2.0, on the other hand, is characterized by synchronicity, interactivity
and the use social networks via mobile devices instead of desktop
computers (\citealp{aghaei2012evolution}). These characteristics
allow users to exploit the Internet to preserve existing relationships,
develop new ties and in some cases promote collective actions, thereby
making online and offline interaction more complements than substitutes.

Other work has assessed how the use of social media correlates with the cognitive dimension of social capital. Results are conflicting, with earlier studies finding a positive relationship in samples of US college students (e.g. \citealp{Ellison_2007}; \citealp{Valenzuela2009}) and more recent studies finding a negative correlation in bigger and representative samples (e.g. \citealp{Sabatini_Sarracino_2017_KYKL}).

The findings of the sociological and communication literature, however,
remain inconclusive due to sample bias, limited observations and the
lack of longitudinal data. Economic studies added important insights
to this literature by reaching more robust conclusions about the relationship
between broadband access and social engagement. The longitudinal
study of \citet{Bauernschuster_2014_JPUBE} did not find evidence
that broadband access displaced offline relationships in post-reunification
Germany. \citet{Falck_Gold_Heblich_2014_AER} and \citet{Gavazza_Nardotto_Valletti_2018_RES},
by contrast, provided evidence that fast Internet crowded out political
participation in Germany and the UK. \citet{Campante_2017_JEEA}
found that the initial spreading of Internet access indeed displaced
political and civic engagement in Italy. In a second moment, however,
the use of online networks triggered new forms of disintermediated
involvement in public affairs that, according to the authors, was
conducive to the later rise in the populist movement.

We add to this field by testing the crowding out hypothesis based
on longitudinal data collected in a representative sample of the British
population, using an identification strategy that has proved promising
for studying the causal effect of broadband penetration, and testing the effect of broadband Internet on a wider range of indicators also including the cognitive dimension of social capital. The United
Kingdom is an interesting case study in that it has an old and large
telephone network, which was designed in the 1930s and irregularly shaped access to fast Internet
in the second half of the 2000s. In addition, the unbundling experience
caused a very rapid broadband penetration in a particularly short
period of time, as it will be explained in detail in Section 3.

\section{The broadband infrastructure in the UK}

\label{sec:Infrastructure} Internet access in the UK between the
end of the 1990s and the first decade of the 2000s mainly relied
on an infrastructure built several decades earlier: the telephone
network. This network has been designed and rolled-out in the 1930s
by the former state monopolist British Telecom (BT), which was formed
as a branch of the UK's General Post Office. The network is made up of
nodes (i.e. the local exchanges LEs), which are connected to each
other to ensure global connectivity, each serving all premises located
in the their respective catchment areas.

The voice network, which was designed for the transmission of the
analog signal of voice communication, could be used to transmit digital
signals, and thus enable all forms of digital communication, among
which Internet access. The main challenge in doing so is given by
the material of telephone lines, which were traditionally made of
copper. A digital signal transmitted on a copper wire suffers
substantial decay with the distance traveled, which made the length
of the copper section of the network a crucial variable determining
local Internet access conditions.

In the 1990s, the copper wires of the network provided low speed connection
to the Internet via dial-up (i.e. a modem connecting to a service
provider by dialing a telephone number). Around 1995 the introduction
of the Digital Subscriber Line technologies (DSL) and later of the
Asymmetric Digital Subscriber Line (ADSL), which use a wider range
of frequencies over the copper line, made it possible to provide Internet
access at a low cost through the voice network. Although the achievable
connection speed by the first versions of ADSL was very limited (and
did not qualify as broadband), technological improvements allowed
reaching a speed of 2Mbit/s (which qualifies as broadband) at the
beginning of the 2000s'. The maximum download speed that could be
reliably provided to users grew rapidly during the first decade of
the 2000s, reaching 24Mbit/s in 2008. The technological possibility
to reach a high connection speed did not however imply that all Internet users
could immediately enjoy such fast access. The adoption of state
of the art DSL technologies required continuous upgrades of the network
and faced a fundamental local limitation: the so-called \emph{last
mile}. As explained, the digital signal suffers strong decay when
it is transmitted on a copper wire, with its strength declining more
then proportionally with the distance traveled. Because of that, BT replaced all connections
between LEs with fiber optic wires, as they account for most of distance
traveled in the communication between the final user and the content
provider/ISP but did not replace all connections between the local
exchanges and the premises.\footnote{The case of fiber until the LE is called \emph{fiber to the cabinet}
(FTTC) while the case of fiber to the premise is \emph{called fiber
to the home} (FTTH). The latter enables faster connections but it
requires also to replace all lines between the LE and the premises and is 
thus much more costly than the former, and almost exclusively
done, at least for the years that we consider in this paper, for large
companies or institutions which had dedicated high bandwidth fiber
connections.} This allowed to reach good Internet speed at a relatively low investment
cost, but generated local differences in the quality of access between
households that depended on the length of their \emph{last mile},
i.e., on the distance between their house and the respective LE. Having
been installed in 1930s for different purposes than the transmission
of the digital signal, the LEs have not been located in order to minimize
the average decay. The LEs' catchment areas are irregularly shaped and
LEs are often not located in their center. Thus, local access conditions
can vary substantially in relatively small areas. In a 2011 report,
Ofcom, the UK's communications regulator says: ``A characteristic
of ADSL broadband is that performance degrades due to signal loss
over the length of the telephone line. This means that the speeds
available to different customers vary significantly, with those with
shorter line lengths (i.e. who live closer to the exchange) typically
able to achieve higher speeds than those with longer line lengths.
{[}...{]} We found that the average download speed received for \emph{up
to} 20 Mbit/s or 24 Mbit/s ADSL packages was 6.6 Mbit/s, and 37\%
of customers had average speeds of 4 Mbit/s or less” (Ofcom, 2011, p. 7).

\begin{figure}[h!]
\centering \includegraphics[width=0.6\textwidth]{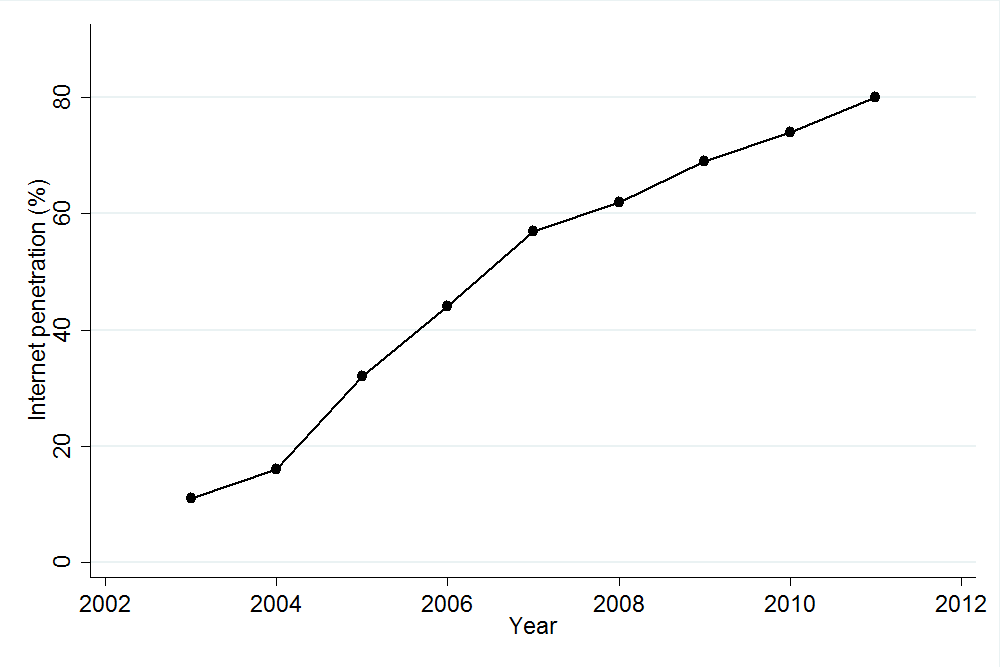}
\caption{Broadband penetration between 2003 and 2011 (source: Eurostat).\label{fig:BBpenOverYear}}
\end{figure}

Broadband Internet penetration in the UK grew slowly between the end of the
1990s and early 2000s (\citealp{deshpande2014examination}). 
This was mainly due to two factors. First, the Internet speed that a private user could obtain
was relatively low, due to the early stage of development of the DSL technology. Second, the institutional setting was not favoring private investments in the sector. The breakthrough in broadband penetration occurred
in 2005, following the implementation of the so-called ``local loop
unbundling” (LLU) required by the European Union's policy on competition
in the telecommunications sector, and introduced in the UK between
2003 and 2004. LLU is the process whereby the incumbent makes its
local voice network available to other companies, and it has been
the cornerstone of the open access legislation in European countries.
Entrants are allowed to put their equipment inside BT's exchanges,
in order to supply customers with an upgrade of their individual voice
lines to DSL services (\citealp{Nardotto_2015_JEEA}). The number of Internet service providers grew rapidly, together with the market share of LLU operators which went from only 2.2\% at the
end of 2005 to almost 40\% at the end of 2009. The process substantially
boosted the broadband penetration in the UK, as shown in Figure~\ref{fig:BBpenOverYear},
which reports the evolution of broadband penetration between 2003
and 2011. In the 5 years from 1999 which can be considered the
starting year of broadband, to 2004 Internet penetration went from
0 to 18\%, while in the following 6 years it almost reached the threshold
of 80\%. 

DSL was not the only option to access the Internet in the years that
we consider, although it was the most popular choice. Until 2010,
approximately 80 percent of broadband accesses in the UK was through
DSL, thereby transforming the old telephone network into the main
determinant of broadband penetration, and the remaining 20 percent
of broadband accesses used the cable network, while less than 0.1
percent relied on fiber and mobile operators.\footnote{The cable network, originally deployed to offer cable-TV, could also
be upgraded to supply Internet access. The cable company, Virgin Media,
made this conversion in parallel to the DSL market and saw its market
share declining from 29\% in 2005 to 22\% in 2010, mainly due to the
increased quality of DSL connections over time (see \citealp{Nardotto_2015_JEEA}
for more details).}

A crucial element in our identification strategy is the length of
the \emph{last mile}. As explained, despite the breakthrough in broadband
penetration, not every household connected to the voice network was
able to access fast Internet with DSL technology. Because of this
technological limitation, the advent of DSL unpredictably turned distance
from the LE into a key determinant of broadband penetration, thereby
creating an exogenous source of variation in access to fast Internet.

\section{Data and empirical strategy}

\label{sec:Data} In this section, we first present the data and we
then detail our identification strategy. In a nutshell, the empirical
analysis exploits individual differences in the actual quality of
Internet access to identify the effect of broadband penetration on
social capital. For this purpose, we combine two sources of data: {[}I{]}
a dataset with detailed geographical information on the topology of
the telephone network provided by Ofcom, the UK Office of Communications,
which we match with {[}II{]} geocoded individual data from the British
Household Panel Survey (BHPS) over the period 1997-2009. The resulting
dataset is a 13 years panel including information on the 
distance of each BHPS respondent from the relative 
node of the telephone
network.


\subsubsection*{ British Household Panel Survey}

The BHPS is a panel survey started in
1991 to address a variety of research topics based on a representative
sample of the British population (\citealp{Taylor2010bhps}) including more than 5,000 households, approximately totaling 10,000
individual interviews. It is household based, and every adult in the
household is interviewed. The BHPS was originally designed as an indefinite
life panel but has now ended. There have been 18 waves of annual interviewing,
with the 18th and last wave completed in 2008. 


In order to match BHPS data with the map of the telephone network provided
by Ofcom, we employ the BHPS Special License\footnote{University of Essex, Institute for Social and Economic Research. (2014). British Household Panel Survey, Waves 1-18, 1991-2009: Special License Access, Lower Layer Super Output Areas and Scottish Data Zones. [data collection]. 3rd Edition. UK Data Service. SN: 6136, \url{ http://doi.org/10.5255/UKDA-SN-6136-2}. 
Application: project number 107760.}
version “Lower Layer Super Output Areas” (LSOA). This upgraded license
provides spatial LSOA references\footnote{LSOAs are the second-narrowest geographical areas in the UK census,
comprising on average 650 households and 1,500 inhabitants. In densely
populated metropolitan areas they correspond to portions of city blocks.} that are crucial for matching the two datasets.

\subsubsection*{Social capital}

Our social capital indicators comprise both the structural and the
cognitive dimensions of the concept as defined by \citet{Uphoff_1999_inbook}.
To measure structural social capital we employ three sets of indicators.
The first set captures the frequency of specific forms of cultural consumption
that are usually enjoyed in company during leisure time, such as watching
movies at the cinema and attending concerts and theatre shows, on a scale ranging
from ``never”, ``once a year or less”, ``several times a year”,
``at least once a month” 
to ``at least once a week”. We transform responses into a dichotomic variable
taking value 0 if the attendance is less than once a month and 1 if
at least once a month.


The second set includes aspects of social connectedness such as the frequency with which respondents meet friends and talk to
neighbors on a scale ranging from ``never”, ``less than once a month”,
``once or twice a month”, ``once or twice a week”, to ``most days”.
We recoded responses into a dichotomic variable equal to 0 if the
frequency is less than once or twice a week and 1 otherwise.

The third set captures political and civic engagement through dummies revealing
whether respondents are members of political parties, trade unions,
professional associations, environmental groups, and other organizations.
The BHPS reports information on membership and active participation
in the form of ``yes/no”\ answers. In our empirical analysis we
consider a respondent as participating in an organization if she is
either a member of it or she declares to participate in its activities.
\footnote{Notice that it is not always the case that a member also declares
that she participates in the organization's activities. On the other hand,
those who declare that they participate in an organization's activities
are not necessarily members.} 

We consider 6 types of organization, which we partition into two
groups based on our elaboration of the literature about the ``Olson-Putnam
controversy''\footnote{In addition to the seminal works of \citet{Olson_1971_Collective_Action}
and \citet{Putnam_1993_Princeton}, see for example \citet{Knack_Keefer_1997_QJE}
and, for a review of the literature, \citet{DegliAntoni_Grimalda_2016_JBEE}.}: Olson-type organizations and Putnam-type organizations. While the former
include political parties, trade unions and professional associations,
the latter includes environmental associations, voluntary service
groups and scout/guides organizations.

Our indicator of cognitive social capital concerns trust towards unknown
others, or ``social trust'', as measured through responses to the
question: ``Generally speaking, would you say that most people can
be trusted, or that you can't be too careful in dealing with people?”
as developed by \citet{Rosenberg_1956_ASR}, possible answers being
``depends”, ``can’t be too careful ”, or ``most people can be trusted”.
We transform responses into a binary variable that takes value 1 if
the answer is that most people can be trusted and 0 otherwise.

In addition to the social capital indicators, we also utilize the
information on the socio-demographic characteristics of respondents,
including age, income, employment status, type of occupation, and
characteristics of the household, the dwelling, and the housing contract, all of which are collected as part of the BHPS. 

\subsubsection*{Broadband infrastructure}

Information on the broadband infrastructure consists of a detailed
map of the topology of the telephone network provided by Ofcom and
previously used in \citet{Nardotto_2015_JEEA} and \citet{ahlfeldt2017speed}.
The data report the  geographic coordinates of each LE and all
the 7-digits post codes served by the node of the network\footnote{There are approximately 1.7 million active post codes in the UK. On
average, a post code covers an area with a radius of 50 meters, but
it is often smaller (i.e., a building) in urban areas.}. Using this information, we are able to reconstruct the exact catchment
area of each LE, and thus compute the (linear) distance between each
household and the respective LE.

\subsection{Empirical strategy}

\label{sec:identification} To identify the causal effect of broadband
Internet on social capital we exploit individual differences in the
actual speed of the connection determined by the variation
in the distance between respondents' dwellings and the respective
LEs, as explained in Section~\ref{sec:Infrastructure}. We estimate
an \emph{intention-to-treat} effect assuming that the penetration
of the broadband resulted in the actual access to fast Internet depending
on the distance from the LE. This approach was used because of the lack of reliable individual-level survey data about the actual access to fast Internet, and because BHPS data on Internet access is available only for a limited number of waves and do not distinguish between fast and slow connections (see Section~\ref{sec:takeup} for further details and for an empirical analysis of how the topology of the network affected the Internet take-up).

Using information on the broadband infrastructure allows us to exploit the panel dimension of the data, as the time
span is long enough to observe the sampled individuals before and
after the introduction of broadband Internet. The panel dimension of the data proves
useful in two respects. First, it allows us to control for unobserved
characteristics that might be correlated both with Internet access
and with our measures of social capital. Second, it allows us to perform
a falsification test by assessing whether distance from the LE is
associated with changes in social capital before the penetration of
the broadband, which further strengthens a causal interpretation of
results. 

We divide the BHPS waves into three periods as reported in Table~\ref{tab:TreatPer} where we mark the waves
in which the indicators employed in the analysis were collected.
We then classify the waves in three periods according to the status
of broadband Internet diffusion. The \textit{Pre-Internet I} period
collects waves between 1997 and 2000, referring to years in which
broadband access was very rare among British households. Years between
2001 and 2004 are in the \textit{Pre-Internet II} period. Just as
in the previous period, broadband access was at most very limited
(overall broadband penetration in 2004 was lower than 20\% and mainly
confined to the richer parts of large cities).
Thus, we consider \textit{Pre-Internet II} as our pre-treatment
condition in the main empirical analysis, and later as our post-placebo
in the falsification test\footnote{The fact that some household could have broadband access already in
this period might bias our results towards zero, and thus our estimates
should be interpreted as lower bounds of the true effect. Instead,
in the case of the falsification test it generates the opposite bias,
with can lead to an overestimate of the true effect and, importantly
of false positives.}.

Finally, the years between 2005 and 2008, when the rapid diffusion
of broadband Internet took place, are in the \textit{Post-Internet}
period. Summarizing, in our main analysis we employ the two most recent
periods: the \textit{Pre-Internet II} period represents the pre-treatment
condition and the \textit{Post-Internet} period is the post-treatment.

In the falsification test, on the other hand we employ the two
periods preceding the diffusion of broadband Internet: the \textit{Pre-Internet
I} period serves as pre-placebo condition while
\textit{Pre-Internet II} now functions as the post-placebo.
\begin{table}[ht!]
\centering \caption{Treatment periods, waves in the BHPS, and questions on social capital.\label{tab:TreatPer}}
\begin{tabular}{lcccc|cccc|cccc}
\hline 
\hline
 

Years:  & \multicolumn{4}{c}{1997-2000} & \multicolumn{4}{c}{2001-2004} & \multicolumn{4}{c}{2005-2008}\\
Period:  & \multicolumn{4}{c}{\textbf{\textit{Pre-Internet I}}} & \multicolumn{4}{c}{\textbf{\textit{Pre-Internet II}}} & \multicolumn{4}{c}{\textbf{\textit{Post-Internet}}}\\

           
& \multicolumn{4}{c}{}         &  \multicolumn{8}{c}{--------------------- \textsc{Main} ---------------------}\\ 
		
				
           & \multicolumn{8}{c}{------------------ \textsc{Placebo} ------------------} &  \multicolumn{4}{c}{\textsc{}}\\


Wave in the BHPS:  & 7  & 8  & 9  & 10  & 11  & 12  & 13  & 14  & 15  & 16  & 17  & 18 \\
\hline 
Leisure activities  &  &  \checkmark  &  &  \checkmark  &  &  \checkmark  &  &  \checkmark  &  &  \checkmark  &  &  \checkmark \\
Organizations  &  \checkmark  &  &  \checkmark  &  &  \checkmark  &  &  \checkmark  &  &  \checkmark  &  &  \checkmark  & \\
Meeting/Talking  & \checkmark  &  \checkmark  &  \checkmark  &  \checkmark  & \checkmark  &  \checkmark  &  \checkmark  &  \checkmark  & \checkmark  & \checkmark  & \checkmark  & \checkmark \\
Trusting & & \checkmark &  & \checkmark &  &  & \checkmark &  & \checkmark &  & \checkmark & \checkmark \\



\hline 
\hline 
\end{tabular}
\begin{spacing}{0.9}
\noindent\begin{minipage}[b]{1\textwidth}%
 \vspace{0.3em}
 {\scriptsize{}Notes: The table reports the division of years into \textit{Pre-Internet I} (from 1997 to2000), \textit{Pre-Internet II} (from 2001 to 2004), and \textit{Post-Internet} (from 2005 to 2008). The table also reports the availability of the different questions in the different waves of the BHPS. \emph{Leisure activities} are the two questions about going to the cinema and going to concerts or theaters. \emph{Organizations} are the questions on being member or an active participant of social organizations. \emph{Meeting/Talking} are the questions on the frequency of meeting friends and talking to neighbors. \emph{Trusting} is the question on the level of trust of other people.} %
\end{minipage}
\end{spacing}
\end{table}

As a result of this partition, and of the heterogeneous availability of the information underlying each of the selected indicators of social capital, the empirical analysis is based on different sub-samples of the full BHPS. The sample selection process leading to each final sub-sample can be described as follows. After selecting the waves of the survey in which the information is available for the selected indicator, separately for \textsc{Main} and \textsc{Placebo} regressions, all observations containing missing values for any of the considered covariates are excluded.

 Individuals are finally kept in the final sub-sample if i) they are observed in at  least one wave of both the relevant \textsc{Pre} and \textsc{Post} period; ii) they do not change LSOA across the considered waves. Importantly, the latter condition avoids the potential bias due to neighborhood sorting on the basis of time-varying unobservable characteristics which may also affect social capital. 

Table ~\ref{tab:SampSel}, in the Appendix, shows the progressive steps of the sample selection process, and their impact on the sample size,  separately for each set of indicators and for main and placebo regressions. The final sub-samples are on average 50\%-60\% the size of the full starting samples containing only the waves of the survey in which information is collected for the relevant set of indicators. Importantly, most of these drops in size are due to the exclusion of individuals who change  their address across waves.

Given that our outcomes are binary variables 
we estimate linear probability models for the effect of the distance
from the LE, determining the quality of Internet access (lower distance
$\longrightarrow$ faster Internet access $\longrightarrow$ higher exposure to the treatment), on our measures of social capital.
Our regression model is reported in~(\ref{eq:mainreg}): 
\begin{equation}
y_{it}=\gamma Distance_{i}\times\textsc{Post}_{t}+\beta X_{it}+Wave_{t}+\eta_{i}+\varepsilon_{it}\label{eq:mainreg}
\end{equation}
where $y_{it}$ is the outcome of interest for person $i$ at year
$t$. $Distance_{i}\times\textsc{Post}_{t}$ is the (reversed) treatment intensity for individual $i$ in year $t$.
This variable is the product of $Distance_{i}$, which is the distance
in kilometers between the individual's house and the LE, and $\textsc{Post}_{t}$,
a binary variable set to 0 if year $t$ is in the \textit{Pre-Internet
II} period, and 1 if year $t$ is in the \textit{Post-Internet}
period. $X_{it}$ is a set of observed time-varying characteristics:
income, household type (categorized), employment status, type of occupation,
age and housing tenure. $\eta_{i}$ is the individual's fixed effect,
which enables us to control for time-invariant unobserved characteristics.
$\varepsilon_{it}$ is the error term.

Controlling for individual fixed effects is crucial for the causal
interpretation of our estimates, as it accounts for the fact that
unobserved (and pre-existent) individual characteristics might be
systematically associated with the propensity for Internet use. As a result, 
our design allows us to isolate the effect of more intense Internet exposure
on social capital. In Section~\ref{sec:takeup}, we will provide
evidence that being located farther away from the LE is associated
both with a lower propensity for having a broadband connection at home
and with less time spent online.

\subsection{Summary statistics}
Table \ref{tab:descstats} reports the descriptive statistics of the
dependent and control variables. To compute the summary statistics
we consider each individual/wave observation that we use in our empirical
analysis. Since the panel is not perfectly balanced, this results in
different sample sizes, depending on the waves considered (see Section~\ref{sec:identification}
for more detail).

{\renewcommand{\arraystretch}{1.12}
\begin{table}[bpth!]
\centering
{\fontsize{10.5pt}{10.5pt}\selectfont
\caption{Descriptive statistics.\label{tab:descstats}}
\input{descstatsfinal}
}
\begin{spacing}{0.9}
\begin{minipage}[b]{1\textwidth}
\vspace{.3em}
\noindent {\scriptsize Notes: Panel A reports the variables that measure social capital. Panel B reports the socio-demographic variables employed in the empirical analysis. In both Panel A and Panel B, the number of observations depends on the BHPS waves considered (see Section~\ref{sec:identification}). In particular, Panel B reports the summary statistics of the socio-demographic variables taking each individual/wave (only once) that enters any of the regression models estimated in Section~\ref{sec:results}.}
\end{minipage}
\end{spacing}
\end{table}
}

Panel A of Table \ref{tab:descstats} reports the summary statistics
of the indicators employed as dependent variables. In our sample, which is representative of the UK's population, 12\% of people regularly go to the cinema, and 5\% attend concerts
and theatre shows. A large majority declares talking regularly with
neighbors and meeting friends, while approximately 40\% believe that
most people can be trusted. Less than 30\% of the population is a member
or participates in the organizations that we consider, with Olson-type
organizations attracting more participation (22\%) than Putnam-type
ones (9\%).

Panel B of Table \ref{tab:descstats} reports summary statistics of
the socio-demographic information collected in the BHPS that we employ
in our empirical analysis.\footnote{Notice that these variables are a subset of the socio-demographic
information collected by the BHPS. We do not employ all possible variables,
but we focus on key demographics (related to age, household composition,
income etc) because in the econometric analysis we employ individual's
fixed effects, and because the panel dimension is relatively short
(at most 6 years).} On average, the BHPS respondent is approximately 50 years old and
earns an annual real income of £27,174. The median individual is employed
or self employed and almost 60\% of households consist of two partners,
with or without children. Most households own their dwelling or have
a mortgage while 20\% rent their habitation privately, or through a housing association or benefiting from
housing assigned by the local government. Finally, the average household
is located 2.7 kilometers from the LE providing Internet access,
with a significant variation, the standard deviation being equal to
2.12 kilometers.

\section{Results}

\label{sec:results} In this section we present the results of the
estimations of model (\ref{eq:mainreg}). We start by analyzing how
the quality of Internet access (decreasing/increasing in the distance from/proximity to the LE)
affects two forms of cultural consumption
that are usually enjoyed in company. We then present results regarding
participation in associations. Finally, we show how access to fast
Internet relates to trust and some forms of offline interaction such
as the frequency with which respondents meet friends and talk to neighbors.

\subsection{Fast Internet and cultural consumption}

\label{sec:ResultLeisure} As discussed in section \ref{sec:Data},
the BHPS does not report either a precise measure of the time spent
in leisure activities, or detailed information on the composition
of leisure time. We thus focus on two forms of cultural consumption
that usually entail social interaction: watching movies at the cinema,
and attending any kind of concert or theater show.

Results are illustrated in Table~\ref{tab:localact}, where columns~(1)
and~(2) report the estimation results for going with the cinema, and
columns~(3) and~(4) for going to concerts or to theaters. In each
pair of regressions we report the results on the main sample (i.e., \textit{Pre-Internet II} and \textit{Post-Internet}) and on the placebo
sample (i.e., \textit{Pre-Internet I} and \textit{Pre-Internet II}). 

\begin{table}[h!]
\centering {\small{}\caption{Effect of fast Internet on leisure activities.\label{tab:localact}}
\input{localact.tex} } 
\end{table}

Results suggest that the faster Internet access associated to a shorter
distance from the LE lowers the time spent on cultural activities.
The size of the effect is economically relevant. The magnitude of
the coefficient implies that one standard deviation reduction (increase)
in the distance from the LE (equal to 2.12 km), causes a reduction
(increase) in the likelihood of going to the cinema of 4.95 percent.

The relationship between broadband access and concerts or theater
attendance is not statistically significant. The falsification
test conducted by assessing how the interaction between the distance
from the LE and the post-placebo relates to our dependent variables
supports a causal interpretation of
the results.

\subsection{Fast Internet and civic engagement}

\label{sec:ResultOrg} We now study the effect of broadband access
on civic engagement, measured as participation in voluntary organizations.
Since its introduction in the pioneering work of \citet{Putnam_1993_Princeton},
membership in organizations is commonly considered as one of the most
significant and reliable indicators of social capital, since it captures
individuals' interest in public affairs and their propensity for contributing
to the public good (see for example \citealp{Knack_Keefer_1997_QJE};
\citealp{Guiso_Sapienza_Zingales_2004_AER}; \citeyear{Guiso_2016_JEEA}).
We define participation as either being a member or actively
participating in an organization's activities. Tables~\ref{tab:dummyorg},
\ref{tab:olson} and \ref{tab:putnam} report estimation results.
In Table~\ref{tab:dummyorg} we partition organizations into two
main types following the social capital literature (e.g. \citealp{Knack_Keefer_1997_QJE}).
Columns~(1)-(2) report results when the outcome variable is participation
in any type of organization, while columns~(3)-(4) and columns~(5)-(6)
respectively refer to participation in Olson- and Putnam-type organizations.
The paired regressions follow the usual order: columns~(1), (3) and~(5)
report the results on the main sample, whereas columns~(2), (4) and~(6)
report the results on the placebo sample.
\begin{table}[h!]
\centering {\small{}\caption{Effect of fast Internet on participation in social organizations.\label{tab:dummyorg}}
\input{dummyorg.tex} } 
\end{table}

Similar to what was found in the previous section, access to
fast Internet (shorter distance from the LE) significantly and strongly
reduces civic engagement. When we consider all the organizations without
distinction, the estimated effect - column (1) - implies that a reduction
(increase) of 1 standard deviation in the distance from the LE, resulting
in a higher connection speed, causes a reduction (increase) in the
likelihood of participation of 3.6 percent.

In Columns~(3) to~(6) we report results for Olson- and Putnam-type
organizations separately. In both cases, the estimated coefficients
are positive and highly statistically significant, suggesting that
participation decreases with the speed of the connection (shorter
distance from the LE). In the case of Olson-type organizations, the
estimates imply that a one-standard deviation increase in the distance
from the LE, resulting in a slower connection, increases the likelihood
of participation by 4 percent. For Putnam-type organizations the estimated
effect is larger and amounting to 6.8 percent.

Finally, we notice that while the estimated coefficients in the main
samples are all statistically significant and display a consistent
pattern, the coefficients estimated on the placebo sample – reported
in columns (2), (4), and (6) – are much smaller in size, do not show
any consistent pattern and are never statistically significant, thus
supporting a causal interpretation of results.

In Tables \ref{tab:olson} and \ref{tab:putnam} we report disaggregated
results for Olson and Putnam organizations. Table~\ref{tab:olson}
focuses on 3 distinct types of Olson-organizations: political parties,
trade unions, and professional organizations. Table~\ref{tab:putnam}
focuses on 3 types of Putnam-organizations: environmental, voluntary
service, and scout organizations. In both tables, we follow the usual
order, reporting the estimates on the main sample in the odd columns
and the estimates on the placebo sample in the even columns.
\begin{table}[h!]
\centering \small{}\caption{Effect of fast Internet on participation in Olson organizations.\label{tab:olson}}
\input{olson.tex} 
\end{table} 

Results reported in Table \ref{tab:olson} show a sizable and statistically
significant effect of access to broadband Internet on participation
in political parties and trade unions. The estimated coefficients
imply that a reduction (increase) of 1 standard deviation in the distance
from the LE, resulting in a faster (slower) connection, causes a reduction
(increase) of the likelihood of participation of 12.7 percent and
4.8 percent respectively.

On the other hand, access to fast Internet does not affect participation
in professional organizations, which is generally aimed at the pursuit
of particular interests and redistributive goals, and mostly takes
place in the context of individuals' professional life, instead of
during their leisure time.

Regarding Putnam organizations, Table~\ref{tab:putnam} reports a
strong and highly statistically significant effect in the case of
scout/guides organizations. The estimated coefficient implies that
a reduction (increase) of 1 standard deviation in the distance from
the LE, resulting in a faster (slower) access to the Internet, causes
a reduction (increase) of the likelihood of participation in this
type of organizations equal to 13.8 percent.
\begin{table}[h!]
\centering {\small{}\caption{Effect of fast Internet on participation in Putnam organizations.\label{tab:putnam}}
\input{putnam.tex} } 
\end{table}

We also find a sizable effect, corresponding to a reduction (increase)
in participation equal to 6 percent for voluntary service organizations,
although in this case the coefficient is significant only at the 10\%
level.

We do not find a strong effect of broadband access on participation
in environmental organizations. However, it is worth noting that,
in the case of Putnam organizations, all coefficients estimated using
the main sample are oriented in the expected direction (while this
is not the case for the coefficients estimated using the placebo sample),
thus indicating a general tendency towards a decrease in participation
with faster Internet access.

Summarizing, we find that better Internet access has a sizable impact
on the engagement in voluntary organizations. This finding holds both
for Olson- and Putnam-type organizations, with some heterogeneity
within these two groups. The estimation results obtained using the
placebo sample support a causal interpretation of the effect.

\subsection{Fast Internet, social interaction and trust}

\label{sec:ResultInter} We conclude our analysis by reporting the
estimates of our empirical model~(\ref{eq:mainreg}) when the outcomes
of interest are related to the frequency of specific forms of social
interaction and the level of trust. (Table~\ref{tab:meettalktrust}).

The frequency of meetings with friends, the habit of talking with neighbors and
social trust seem to be unaffected by broadband Internet, suggesting
that the displacement effect does not take place in these cases. The
null result regarding the frequency of meetings may stem from the
opposing effects that the literature has attributed to Internet use
in the development of offline relationships. While several authors
have highlighted the risk that ICTs may crowd out interaction with
friends and relatives (e.g. \citealp{Putnam_2000_Bowling_Alone};
\citealp{Nie_et_al_2002_}; \citealp{Gershuny_2003_SFORCES}), more
recent studies suggest that access to fast Internet may also be a
tool for preserving and developing existing relationships despite
time and distance constraints, thank to the interactivity of social
networking sites (SNS) (\citealp{Ellison_2007}; \citealp{Bauernschuster_2014_JPUBE};
\citealp{Antoci_et_al_2015_JMAS}).
\begin{table}[h!]
\centering {\small{}\caption{Effect of fast Internet on social interactions and trust.\label{tab:meettalktrust}}
\input{meettalktrust.tex} } 
\end{table}

Chats with neighbors, on the other hand, generally occur occasionally
and incidentally, for example when leaving or coming back home. They
thus seem to be less or not at all vulnerable to the displacement
effect possibly caused by Internet use.

Social trust is a cognitive phenomenon depending on individuals' values
and perceptions, unrelated to time constraints and less sensitive
to the risk of crowding out. Some studies suggested that Internet
use may be detrimental to trust to the extent to which it entails
engagement in interactions with strangers on SNS, due to the phenomenon
of online incivility recently spreading on platforms such as Facebook
and Twitter (\citealp{Antoci_et_al_2016_PLOS}; \citealp{Sabatini_Sarracino_2017_KYKL}).
Our data, however, refers to an earlier period in which social media
just started flourishing and online incivility was a much rarer
phenomenon (\citealp{Rost_Stahel_Frey_2016_PLOS}).


\section{Effect of distance on Internet take-up}

\label{sec:takeup} The empirical analysis carried out in sections~\ref{sec:Data}
and \ref{sec:results} relies on the fact that, in the years we consider,
the distance between the geographical location of the households and
the respective LE was crucial in affecting the actual quality of broadband
Internet access. The use of technological factors
as a source of exogenous variation in Internet access conditions is supported by 1) technical reports produced by the
industry and the regulator, which emphasize the role of distance as a crucial determinant of connection speed;
(\citealp{Ofcom2010}; \citealp{Ofcom2011}; \citealp{Ofcom2012};
\citealp{Ofcom2013}); and 2) other studies that also
make use of factors affecting the quality of Internet access (such as the distance of the dwellings from the LE) to identify the effect of broadband penetration on a range of hypothetical outcomes. 
 (see, for instance, \citealp{Amaral_et_al_2017_health}; \citealp{Falck_Gold_Heblich_2014_AER}).




Most of those studies took the effect of distance between the premises and the LEs on the Internet take up for granted, although, due to the lack of suitable data, they were not able
to test it empirically (e.g. \citealp{ahlfeldt2017speed}; \citealp{Falck_Gold_Heblich_2014_AER}).
In this section we use self-reported information on individual Internet
access collected in the BHPS to provide evidence that distance from
the LE actually affected the access to broadband Internet in the UK. 


For this purpose, we use a question asked in some waves of the BHPS
on broadband Internet access. Respondents were asked whether they
had a fixed-line broadband connection at home at the time of the interview
and, in the affirmative case, how much time they spent online on average
per day. 
On one hand, answers to these questions are worth examining because
they can provide evidence on the effect of distance on take up. On
the other hand, they have certain limitations, which is why we do
not exploit this information in the main empirical analysis. The first
limitation is that this variable is likely to be subject to
measurement error. Most users hardly knew whether a connection could be defined as broadband and what the broadband
speed/technology actually was, with some of them answering affirmatively when they actually had a slow DSL connection (such as an ISDN) or an only in theory broadband connection that actually suffered from substantial decay due to the distance from the LE.\footnote{In our period of analysis, broadband packages were typically advertised referring to their \emph{theoretical} maximum speed while information about the actual or average speed was omitted by providers. As a result, consumers largely ignored the fact that the actual speed depended by the distance from LEs. According to Ofcom reports focusing on consumer protection issues, this has been a source of major confusion for UK Internet users (\citealp{ofcom2006}; \citealp{ofcom2009short}). The remarkable 60\% gap between advertised and actual speed registered in the UK was far above the EU average of 40\% (\citealp{eu2012}). Still in 2016,  an Ofcom research found that even ``business consumers --particularly small or medium sized enterprises-- are confused about how the \textit{actual} speed of their broadband service compared to the \textit{headline} maximum speed used in advertising'' (\citealp{ofcom2016}; \citealp{hc2017}). For these reasons, consumer self-reported survey data on Internet connections are considered not completely reliable  (most likely upward biased) and therefore only weakly informative (\citealp{oecd2008}).} In addition, respondents might give a positive answer even when the connection was not at home (which is crucial for instance in the allocation of leisure time) but for example at the workplace.

The second limitation is due to the waves in which this information
was collected and the sample of households that were interviewed,
which is not perfectly consistent with the other samples employed
in the rest of the empirical analysis. Although these limitations prevented us from using information on individual broadband access in a TSLS strategy, it is worth investigating the link between distance and Internet access, keeping in mind that the following results are not fully conclusive.

The empirical model we employ, reported in equation~(\ref{eq:takeup}), is similar to~(\ref{eq:mainreg}) that has been employed in previous sections.
\begin{equation}
	Internet_{it}=\gamma Distance_{i}+\beta X_{it}+Wave_{t}+\eta_{i}+\varepsilon _{it}
\label{eq:takeup}
\end{equation}
Here, the outcome $Internet_{it}$ is in one case the presence of a broadband connection at home, and in the other case the time spent online; $Distance_{i}$ is the distance between the household and the respective LE, and, as in previous models, $X_{it}$ is a set of time-varying controls including income, household type (categorized), employment status, type of occupation, age and tenure; $\eta_i$ is the individual's random effect; and $\varepsilon _{it}$ is the error term.

The main difference with respect to model~(\ref{eq:mainreg}) is that we cannot employ a fixed-effect estimator, due to the fact that our main variable of interest, the distance between the home and the LE, is constant over time and because the question is asked only in the last waves of the BHPS, all belonging to the \textit{Post-Internet} period. Hence, we employ a random-effect estimator (with controls) when the dependent variable is the presence of broadband connection at home, and OLS estimator (with controls) when the dependent variable is the time spent online.

\begin{figure}[ht!]
	\centering
		\includegraphics[width=0.90\textwidth]{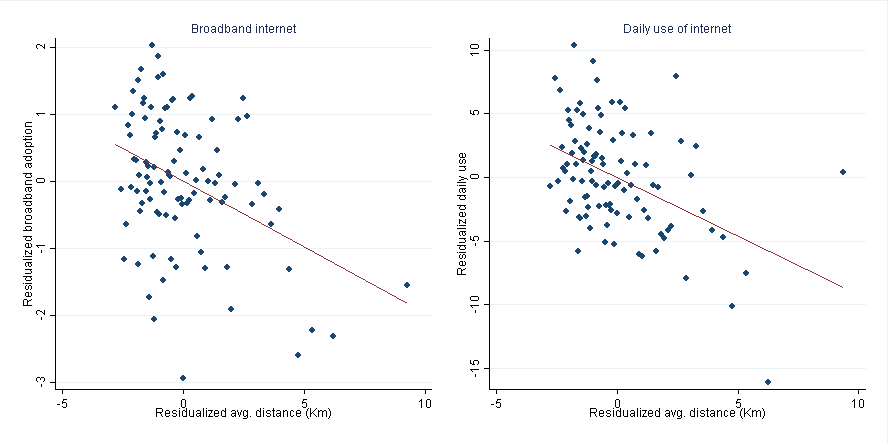}
		
	\caption{Internet access and distance between the house and the LE. Both panels report on the x-axis the residuals of a regression of the distance between the house and the respective LE on the socio-demographic control variables. The y-axis of the left panel reports the residual of a regression for the presence of broadband at home on the socio-demographic control variables. The y-axis of the rigth panel reports the residual of a regression for the time spent online on the socio-demographic control variables.}
	\label{fig:takeup}
\end{figure}

A clear negative relationship between the distance and broadband access emerges from the data, and it holds both for the presence of a broadband connection at home (the extensive margin), and for the time spent online (the intensive margin). This negative relationship can be appreciated in Figure~\ref{fig:takeup}, that results from the estimation of our model, and a straightforward application of the Frisch-Waugh-Lowell theorem (see, for instance \citealt{davidson2004econometric}), that enables us to partial-out the role of socio-demographics when we examine the relation between take up and distance. We split the set of regressors into two groups: the geographical distance between the households and the respective LEs, and the socio-demographic control variables. We report on the y-axis the residuals of model~\ref{eq:takeup} where we employ as independent variables only the socio-demographic control variables (i.e., we exclude the $Distance$), and on the x-axis the residuals of a regression of $Distance$ on the socio-demographic control variables. Residuals are grouped in 100 bins for which we take the average value.

As explained, Figure~\ref{fig:takeup} is strictly related to the main model in equation~(\ref{eq:takeup}), whose estimates are reported in Table~\ref{tab:takeup}.
\begin{table}[ht!]
\centering
{\small
\caption{Effect of distance from the LE on fast Internet take up.\label{tab:takeup}}
\input{takeup}
}
\end{table}
Column (1) of the table shows the estimated effect of distance between the house and the respective LE on the adoption of a broadband connection. The coefficient, as expected, is negative and statically significant, thus indicating that distance has an influence on Internet access and the \emph{extensive margin}. Furthermore, column (2) of the table reports the estimated effect of distance on the time spent online, again showing a strong, negative impact of distance on the access, which in this case is a measure of the \emph{intensive margin}. Thus, we can conclude that the distance, by affecting the quality and the reliability of the connection, the running costs for the Internet providers, and in turn possibly the availability (as pointed out in \citealp{Falck_Gold_Heblich_2014_AER}, given the status of the technology, Internet quality could be so low beyond 4.2 kilometers that broadband connections until 2008 could simply be not available), 
was a crucial element of the fast Internet take up up and of the time that users spent online.



\section{Conclusions}

In this paper we studied how the penetration of broadband Internet
affected social capital in the UK. Matching unique information on
the topology of the old voice communication infrastructure with geocoded
survey data on individual behaviors, we were able to exploit discontinuities
in the speed of Internet connection to test whether the availability
of fast Internet displaced offline activities such as civic engagement,
political participation, and other forms of face-to-face interaction.
Our results paint a complex picture. As in \citet{Bauernschuster_2014_JPUBE},
we do not find evidence that broadband access displaced offline relationships
such as meetings with friends. 
On the other hand, the empirical analysis in this paper shows that fast Internet crowded out forms of cultural consumption that are usually enjoyed in company such as watching movies at the cinema and attending concerts and theatre shows. In addition, broadband penetration significantly displaced civic engagement and political participation, i.e. time consuming activities that usually take place during leisure time, are not pursued in order to reach particularistic goals, and generally relate to a non self-interested involvement in public affairs.
Associational activities have been often mentioned as forms of bridging social capital
creating positive societal and economic externalities (\citealp{Putnam_1993_Princeton};
\citeyear{Putnam_1995_JD}), which has recently started to decline in many
OECD countries (e.g. \citealp{Putnam_2000_Bowling_Alone}; \citealp{Costa_Kahn_2003_Kyklos}).
In this respect, our study offers a possible interpretation of the
``decline in community life-thesis'' (\citealp{Paxton_1999_AJS})
supporting Putnam's early intuitions on the detrimental role of technological
progress in social cohesion. The picture provided by our findings
is also consistent with the patterns of declining engagement in public
affairs previously sketched by those Internet studies in economics
that analyzed how the first stage of broadband penetration affected
political participation before the advent of social media (\citealp{Falck_Gold_Heblich_2014_AER};
\citealp{Campante_2017_JEEA}; \citealp{Gavazza_Nardotto_Valletti_2018_RES}). 

The developing role of fast Internet use, however, certainly calls
for further investigation, as new forms of SNS-mediated civic and
political participation spread after the period covered by our data.
A more recent wave of Internet studies suggests that social media
may also support collective action and political mobilization, especially
in young democracies and authoritarian regimes, thereby providing
a potentially positive contribution to the strengthening of political participation
(\citealp{Enikolopv_et_al_2016_CEPR};
\citealp{Enikolopv_et_al_2016_AEJ}). 
Other studies, on the other
hand, highlight how the increasing importance of social media in the
public discourse entails 
new systemic risks connected to the propagation of misinformation 
(\citealp{DelVicario_et_al_2016_PNAS}), the extreme polarization of the political debate (\citealp{Muller_Schwartz_2018_SSRN}) and the spreading of online incivility (\citealp{antoci2018civility})
Future research should deal with these conflicting effects, also in
light of the prominent role that a limited number of platforms, such
as Facebook and Twitter, assumed in biasing electoral results (\citealp{Allcot_Gentzkow_2017_JEP}).


\clearpage{}\newpage{} \bibliographystyle{chicago} 
\bibliography{bibliography2017}


\clearpage

\section*{Appendix}
\addcontentsline{toc}{section}{Appendix}
\setcounter{subsection}{0} 		\renewcommand{\thesubsection}{A.\arabic{subsection}}
\setcounter{subsubsection}{0} 	\renewcommand{\thesubsubsection}{A.1.\arabic{subsubsection}}


\begin{table}[h!]
\small
\centering \caption{Sample selection steps.\label{tab:SampSel}}
\begin{tabular}{lcccccccc} \hline\hline  & \multicolumn{2}{c}{Leisure activities} & \multicolumn{2}{c}{Organizations}   & \multicolumn{2}{c}{Meeting-Talking} & \multicolumn{2}{c}{Trusting}\\  \cmidrule(l{4pt}r{4pt}){2-3}\cmidrule(l{4pt}r{4pt}){4-5}\cmidrule(l{4pt}r{4pt}){6-7}\cmidrule(l{4pt}r{4pt}){8-9} &    \multicolumn{1}{c}{\textsc{Main}} & \multicolumn{1}{c}{\textsc{Placebo}} &    \multicolumn{1}{c}{\textsc{Main}} & \multicolumn{1}{c}{\textsc{Placebo}} &    \multicolumn{1}{c}{\textsc{Main}} & \multicolumn{1}{c}{\textsc{Placebo}} &    \multicolumn{1}{c}{\textsc{Main}} & \multicolumn{1}{c}{\textsc{Placebo}}\\

\multicolumn{9}{l}{}\\
\multicolumn{9}{l}{1 - Starting samples containing the relevant waves only}\\
\hline
PIDs             &       20310&       22251&       22601&       23241&       23760&       24648&       19318&       20933\\
Obs.             &      60748&       61777&       63889&       59825&      124637&      116944&       59631&       41329\\
\multicolumn{9}{l}{}\\
\multicolumn{9}{l}{2 - Ecluding observations with missing values for any covariate}\\
\hline
PIDs             &       19207&       21313&       21489&       22396&       22612&       23766&       17684&       17370\\
Obs.       &      55952&       57455&       59155&       56500&      115133&      110067&       51099&       36072\\
\multicolumn{9}{l}{}\\
\multicolumn{9}{l}{3 - Excluding PIDs not oberved at least once in both PRE and POST period}\\
\hline
PIDs              &      12884&       13720&       13050&       12538&       13845&       13825&       10208&       10266\\
Obs.        &       47513&       47773&       48209&       42240&       98608&       86971&       37161&       26642\\
\multicolumn{9}{l}{}\\
\multicolumn{9}{l}{4 - Excluding individuals who change LSOA across the considered waves}\\
\hline
PIDs             &         9670&       10397&        9616&        9068&        9800&        9631&        7863&        7740\\
Obs.          &       35697&       35860&       35656&       30193&       70194&       59572&       28528&       19827\\
\hline\hline
\end{tabular}

\begin{spacing}{0.9}
\noindent\begin{minipage}[b]{1\textwidth}%
 \vspace{0.3em}
 {\scriptsize{}Notes: The table reports the number of PIDs and observations left out after each step of the sample selection process. Figures are reported separately for each set of social capital indicators, and for \textsc{Main} and \textsc{Placebo} regressions } %
\end{minipage}
\end{spacing}

\end{table}

\end{document}

%% file: descstatsfinal.tex
\begin{tabular}{lccccc}\hline\hline \\
\textbf{Panel A: Dependent Variables} & Mean & St.Dev. & Min & Max & Obs \\
\cline{2-6}
Cinema attendance        &        0.12&        0.33&        0.00&        1.00&       35697\\
Theatre and concerts attendance &        0.05&        0.21&        0.00&        1.00&       35698\\
Any organization    &        0.28&        0.45&        0.00&        1.00&       35656\\
Olson organizations &        0.22&        0.42&        0.00&        1.00&       35656\\
\hspace{0.1cm} - Political party     &        0.02&        0.15&        0.00&        1.00&       35656\\
\hspace{0.1cm} - Trade union         &        0.15&        0.36&        0.00&        1.00&       35656\\
\hspace{0.1cm} - Professional organizations&        0.08&        0.27&        0.00&        1.00&       35656\\
Putnam organizations&        0.09&        0.29&        0.00&        1.00&       35656\\
\hspace{0.1cm} - Environmental organizations&        0.03&        0.17&        0.00&        1.00&       35656\\
\hspace{0.1cm} - Voluntary organizations&        0.05&        0.21&        0.00&        1.00&       35656\\
\hspace{0.1cm} - Scout organizations &        0.02&        0.14&        0.00&        1.00&       35656\\
Talking to neighbors   &        0.80&        0.40&        0.00&        1.00&       70194\\

Meetings with friends         &        0.87&        0.34&        0.00&        1.00&       70189\\
Most people can be trusted        &        0.40&        0.49&        0.00&        1.00&       28528\smallskip \\
\hline
\textbf{Panel B: Demographic characteristics} \\
Age                 &       48.62&       18.11&       15.00&      101.00&       84946\\
Log of real household income&       10.22&        0.75&       -0.55&       14.14&       84950\\
Household type:\\
\hspace{0.1cm} - Single non elderly  &        0.06&        0.23&        0.00&        1.00&       84950\\
\hspace{0.1cm} - Single elderly      &        0.09&        0.28&        0.00&        1.00&       84950\\
\hspace{0.1cm} - Couple: no children &        0.30&        0.46&        0.00&        1.00&       84950\\
\hspace{0.1cm} - Couple: dependent children&        0.30&        0.46&        0.00&        1.00&       84950\\
\hspace{0.1cm} - Couple: non dependent children&        0.13&        0.34&        0.00&        1.00&       84950\\
\hspace{0.1cm} - Lone parent: dependent children&        0.05&        0.21&        0.00&        1.00&       84950\\
\hspace{0.1cm} - Lone parent: non dependent children&        0.04&        0.20&        0.00&        1.00&       84950\\
\hspace{0.1cm} - 2+ unrelated adults &        0.01&        0.09&        0.00&        1.00&       84950\\
\hspace{0.1cm} - Other               &        0.02&        0.13&        0.00&        1.00&       84950\\
Employment status:\\
\hspace{0.1cm} - Employed or self-employed&        0.55&        0.50&        0.00&        1.00&       84950\\
\hspace{0.1cm} - Unemployed          &        0.03&        0.17&        0.00&        1.00&       84950\\
\hspace{0.1cm} - Inactive            &        0.42&        0.49&        0.00&        1.00&       84950\\
Type of occupation:\\
\hspace{0.1cm} - Higher managerial and professional occupations&        0.06&        0.24&        0.00&        1.00&       84950\\
\hspace{0.1cm} - Lower managerial and professional occupations&        0.15&        0.36&        0.00&        1.00&       84950\\
\hspace{0.1cm} - Intermediate occupations (clerical, sales, service)&        0.08&        0.28&        0.00&        1.00&       84950\\
\hspace{0.1cm} - Small employers and own account workers&        0.06&        0.23&        0.00&        1.00&       84950\\
\hspace{0.1cm} - Lower supervisory and technical occupations&        0.05&        0.23&        0.00&        1.00&       84950\\
\hspace{0.1cm} - Semi-routine occupations&        0.11&        0.31&        0.00&        1.00&       84950\\
\hspace{0.1cm} - Routine occupations &        0.07&        0.26&        0.00&        1.00&       84950\\
\hspace{0.1cm} - Unemployed/Inactive &        0.41&        0.49&        0.00&        1.00&       84950\\
Housing tenure:\\
\hspace{0.1cm} - Owned outright      &        0.34&        0.47&        0.00&        1.00&       84950\\
\hspace{0.1cm} - Owned with mortgage &        0.45&        0.50&        0.00&        1.00&       84950\\
\hspace{0.1cm} - Rented privately    &        0.05&        0.22&        0.00&        1.00&       84950\\
\hspace{0.1cm} - Rented - housing assoc. or local authority&        0.16&        0.36&        0.00&        1.00&       84950\\
Distance house-LE (km)       &        2.70&        2.11&        0.01&       24.33&       84950\\
\hline\hline\end{tabular}

%% file: localact.tex
\begin{tabular}{lcccc} \hline\hline Dep. Variables: & \multicolumn{2}{c}{Cinema attendance} & \multicolumn{2}{c}{Theatre and concerts } \\ & \multicolumn{2}{c}{} & \multicolumn{2}{c}{attendance} \\ \cmidrule(l{4pt}r{4pt}){2-3}\cmidrule(l{4pt}r{4pt}){4-5} & \multicolumn{1}{c}{\textsc{Main}} & \multicolumn{1}{c}{\textsc{Placebo}} & \multicolumn{1}{c}{\textsc{Main}} & \multicolumn{1}{c}{\textsc{Placebo}}\\ & (1) & (2) & (3) & (4) \\ \cmidrule(l{4pt}r{4pt}){2-3}\cmidrule(l{4pt}r{4pt}){4-5}
Distance $\times$ Post&        0.28** &       -0.17   &        0.09   &       -0.13   \\
                    &      (0.12)   &      (0.13)   &      (0.09)   &      (0.10)   \\
Controls & \checkmark & \checkmark & \checkmark & \checkmark \\ Time FEs & \checkmark & \checkmark & \checkmark & \checkmark \\ Individual FEs & \checkmark & \checkmark & \checkmark & \checkmark \\\hline
Observations        &       35695   &       35859   &       35696   &       35864   \\
R\textsuperscript{2}&        0.01   &        0.01   &        0.00   &        0.00   \\
Num. PIDs           &        9670   &       10397   &        9670   &       10397   \\
\hline\hline\end{tabular}

%% file: dummyorg.tex
\begin{tabular}{lcccccc} \hline\hline Dep. Variables: & \multicolumn{2}{c}{Any organization} & \multicolumn{2}{c}{Olson organizations} & \multicolumn{2}{c}{Putnam organizations}\\ \cmidrule(l{4pt}r{4pt}){2-3}\cmidrule(l{4pt}r{4pt}){4-5}\cmidrule(l{4pt}r{4pt}){6-7} & \multicolumn{1}{c}{\textsc{Main}} & \multicolumn{1}{c}{\textsc{Placebo}} & \multicolumn{1}{c}{\textsc{Main}} & \multicolumn{1}{c}{\textsc{Placebo}} & \multicolumn{1}{c}{\textsc{Main}} & \multicolumn{1}{c}{\textsc{Placebo}}\\ & (1) & (2) & (3) & (4)& (5) & (6) \\ \cmidrule(l{4pt}r{4pt}){2-3}\cmidrule(l{4pt}r{4pt}){4-5}\cmidrule(l{4pt}r{4pt}){6-7}
Distance $\times$ Post&        0.48***&       -0.09   &        0.42***&       -0.08   &        0.29***&        0.01   \\
                    &      (0.14)   &      (0.18)   &      (0.13)   &      (0.16)   &      (0.10)   &      (0.14)   \\
Controls & \checkmark & \checkmark & \checkmark & \checkmark & \checkmark & \checkmark \\ Time FEs & \checkmark & \checkmark & \checkmark & \checkmark & \checkmark & \checkmark \\ Individual FEs & \checkmark & \checkmark & \checkmark & \checkmark & \checkmark & \checkmark \\\hline
Observations        &       35654   &       30192   &       35654   &       30192   &       35654   &       30192   \\
R\textsuperscript{2}&        0.01   &        0.01   &        0.02   &        0.02   &        0.00   &        0.00   \\
Num. PIDs           &        9616   &        9068   &        9616   &        9068   &        9616   &        9068   \\
\hline\hline\end{tabular}

%% file: olson.tex
\begin{tabular}{lcccccc} \hline\hline Dep. Variables: & \multicolumn{2}{c}{Political parties} & \multicolumn{2}{c}{Trade unions} & \multicolumn{2}{c}{Professional}\\ & \multicolumn{2}{c}{} & \multicolumn{2}{c}{} & \multicolumn{2}{c}{organizations}\\ \cmidrule(l{4pt}r{4pt}){2-3}\cmidrule(l{4pt}r{4pt}){4-5}\cmidrule(l{4pt}r{4pt}){6-7} & \multicolumn{1}{c}{\textsc{Main}} & \multicolumn{1}{c}{\textsc{Placebo}} & \multicolumn{1}{c}{\textsc{Main}} & \multicolumn{1}{c}{\textsc{Placebo}} & \multicolumn{1}{c}{\textsc{Main}} & \multicolumn{1}{c}{\textsc{Placebo}}\\ & (1) & (2) & (3) & (4)& (5) & (6) \\ \cmidrule(l{4pt}r{4pt}){2-3}\cmidrule(l{4pt}r{4pt}){4-5}\cmidrule(l{4pt}r{4pt}){6-7}
Distance $\times$ Post&        0.12** &        0.03   &        0.34***&       -0.05   &       -0.07   &        0.16   \\
                    &      (0.05)   &      (0.06)   &      (0.11)   &      (0.13)   &      (0.09)   &      (0.11)   \\
Controls & \checkmark & \checkmark & \checkmark & \checkmark & \checkmark & \checkmark \\ Time FEs & \checkmark & \checkmark & \checkmark & \checkmark & \checkmark & \checkmark \\ Individual FEs & \checkmark & \checkmark & \checkmark & \checkmark & \checkmark & \checkmark \\\hline
Observations        &       35654   &       30192   &       35654   &       30192   &       35654   &       30192   \\
R\textsuperscript{2}&        0.00   &        0.00   &        0.02   &        0.02   &        0.00   &        0.01   \\
Num. PIDs           &        9616   &        9068   &        9616   &        9068   &        9616   &        9068   \\
\hline\hline\end{tabular}

%% file: putnam.tex
\begin{tabular}{lcccccc} \hline\hline Dep. Variables: & \multicolumn{2}{c}{Environmental} & \multicolumn{2}{c}{Voluntary service} & \multicolumn{2}{c}{Scout}\\ & \multicolumn{2}{c}{organizations} & \multicolumn{2}{c}{organizations} & \multicolumn{2}{c}{organizations}\\ \cmidrule(l{4pt}r{4pt}){2-3}\cmidrule(l{4pt}r{4pt}){4-5}\cmidrule(l{4pt}r{4pt}){6-7} & \multicolumn{1}{c}{\textsc{Main}} & \multicolumn{1}{c}{\textsc{Placebo}} & \multicolumn{1}{c}{\textsc{Main}} & \multicolumn{1}{c}{\textsc{Placebo}} & \multicolumn{1}{c}{\textsc{Main}} & \multicolumn{1}{c}{\textsc{Placebo}}\\ & (1) & (2) & (3) & (4)& (5) & (6) \\ \cmidrule(l{4pt}r{4pt}){2-3}\cmidrule(l{4pt}r{4pt}){4-5}\cmidrule(l{4pt}r{4pt}){6-7}
Distance $\times$ Post&        0.09   &        0.04   &        0.14*  &        0.05   &        0.13***&       -0.05   \\
                    &      (0.06)   &      (0.08)   &      (0.08)   &      (0.11)   &      (0.05)   &      (0.07)   \\
Controls & \checkmark & \checkmark & \checkmark & \checkmark & \checkmark & \checkmark \\ Time FEs & \checkmark & \checkmark & \checkmark & \checkmark & \checkmark & \checkmark \\ Individual FEs & \checkmark & \checkmark & \checkmark & \checkmark & \checkmark & \checkmark \\\hline
Observations        &       35654   &       30192   &       35654   &       30192   &       35654   &       30192   \\
R\textsuperscript{2}&        0.00   &        0.00   &        0.00   &        0.00   &        0.00   &        0.00   \\
Num. PIDs           &        9616   &        9068   &        9616   &        9068   &        9616   &        9068   \\
\hline\hline\end{tabular}

%% file: meettalktrust.tex
\begin{tabular}{lcccccc} \hline\hline Dep. Variables: & \multicolumn{2}{c}{Talking to neighbors} & \multicolumn{2}{c}{Meetings with friends} & \multicolumn{2}{c}{Trusting people} \\ \cmidrule(l{4pt}r{4pt}){2-3}\cmidrule(l{4pt}r{4pt}){4-5}\cmidrule(l{4pt}r{4pt}){6-7} & \multicolumn{1}{c}{\textsc{Main}} & \multicolumn{1}{c}{\textsc{Placebo}} & \multicolumn{1}{c}{\textsc{Main}} & \multicolumn{1}{c}{\textsc{Placebo}} & \multicolumn{1}{c}{\textsc{Main}} & \multicolumn{1}{c}{\textsc{Placebo}}\\ & (1) & (2) & (3) & (4)& (5) & (6) \\ \cmidrule(l{4pt}r{4pt}){2-3}\cmidrule(l{4pt}r{4pt}){4-5}\cmidrule(l{4pt}r{4pt}){6-7}
Distance $\times$ Post&        0.01   &       -0.04   &       -0.09   &        0.10   &       -0.36   &       -0.10   \\
                    &      (0.11)   &      (0.13)   &      (0.10)   &      (0.13)   &      (0.24)   &      (0.28)   \\
Controls & \checkmark & \checkmark & \checkmark & \checkmark & \checkmark & \checkmark \\ Time FEs & \checkmark & \checkmark & \checkmark & \checkmark & \checkmark & \checkmark \\ Individual FEs & \checkmark & \checkmark & \checkmark & \checkmark & \checkmark & \checkmark \\\hline
Observations        &       70190   &       59571   &       70185   &       59570   &       28525   &       19827   \\
R\textsuperscript{2}&        0.00   &        0.01   &        0.00   &        0.00   &        0.03   &        0.01   \\
Num. PIDs           &        9800   &        9631   &        9799   &        9632   &        7863   &        7740   \\
\hline\hline\end{tabular}

%% file: takeup.tex
\begin{tabular}{lcc} \hline\hline Dep. Variables: & \multicolumn{1}{c}{Broadband internet} & \multicolumn{1}{c}{Daily use of internet} \\             & (1) & (2) \\ \cmidrule(l{4pt}r{4pt}){2-2} \cmidrule(l{4pt}r{4pt}){3-3}
Distance            &       -0.89***&       -0.90***\\
                    &      (0.23)   &      (0.22)   \\
Controls                & \checkmark & \checkmark \\                 Time FEs                & \checkmark & no \\                 Individual REs  & \checkmark & no \\\hline
Observations        &       36081   &       13066   \\
R\textsuperscript{2}&        0.30   &         0.18      \\
Num. PIDs           &       14064   &         13066      \\
\hline\hline\end{tabular}